\documentclass[preprint,showpacs,showkeywords,preprintnumbers,amsmath,amssymb]{revtex4}
\usepackage{graphicx}
\usepackage{bm}
\begin{document}

\title{Time-dependent barrier passage in External Noise Modulated System-reservoir Environment}

\author{WANG Chun-Yang}
\thanks{Corresponding author. Email: wchy@mail.bnu.edu.cn}
\author{HUANG Jun}
\affiliation{Department of Physics and engineering, Qufu Normal
University, Qufu, 273165, China}


\begin{abstract}
The time-dependent barrier passage of an anomalous system-reservoir
coupling non-equilibrium open environment is studied where the heat
bath is modulated by an external noise. The time-dependent barrier
passing probability is obtained analytically by solving the
generalized Langevin equation. The escaping rate and effective
transmission coefficient are calculated by using of the reactive
flux method in the particular case of external $\delta$-correlated
noise modulated internal Ornstein-Uhlenbeck process. It is found
that not all the cases of external noise modulating is harmful to
the rate process. Sometimes it is even beneficial to the diffusion
of the particle. There exists an optimal strength of external noise
modulation for the particle to obtain a biggest probability to
escape from the potential well to form a notable rate of effective
net flux.
\end{abstract}

\pacs{47.70.-n, 82.20.Db, 82.60.-s, 05.60.Cd}

\maketitle

\section{\label{sec:level1}INTRODUCTION}

The diffusion model for chemical reactions has become ubiquitous in
many areas of physics, chemistry and biology since its advancing by
H. A. Kramers in terms of the theory of Brownian motion in phase
space \cite{kramers}. Several of its variants have been proposed for
understanding the nature of activated processes in classical
\cite{kampen,groteh,peterh,epolla}, semiclassical \cite{ddsray}, and
quantum systems \cite{clmode,glingo,grabet,peteha}. The majority of
these treatments concern essentially with an equilibrium thermal
bath at a finite temperature that simulates the reaction coordinates
to cross the activation energy barrier and the inherent noise of the
medium originates internally. Therefore these systems are actually
classified as a thermodynamically closed system-plus-reservoir
environment in contrast to the systems directly driven by external
noise(s) in non-equilibrium statistical mechanics \cite{met1}.
However, what will happen if the total system-reservoir environment
is really open or modulated by some external forces, is the barrier
crossing dynamics still amenable to the present theoretical analysis
? These problems are of valuable consideration from a microscopic
point of view.

Recently, A type of system-reservoir coupling model simulating an
activated rate process is studied where the heat bath is modulated
by an external noise \cite{dsray2,dsray3}. A non-equilibrium
stationary state distribution which is reminiscent of the
equilibrium Boltzmann distribution in calculation of rate is
proposed for this open environment by analytically solving the
generalized Fokker-Planck equation and a generalized Kramers' rate
is derived via the method of flux-over-population \cite{fopm}. This
brings a new broad perspective for the study of anomalous activated
processes in non-equilibrium open systems. However, some crucial
problems such as the barrier recrossing phenomenon which is always
met in an escaping process have not been concerned in these works.
The time-dependent dynamical barrier escaping process in
non-equilibrium open systems is far from well studied. We noticed
that these can be easily achieved by the method of reactive flux
\cite{rf1,rf2,jcp} which is developed from the transition state
theory \cite{TST1,TST2,TST3}. Therefore, following this powerful and
convenient way we present in this paper a careful study on the
non-equilibrium activated barrier crossing process concerning
intensively on its dynamical details.

The paper is organized as follows: In Sec. II, the expression of the
barrier passing probability is obtained by analytically solving the
relevant generalized Langevin equation. In Sec. III, we give the
time-dependent barrier escaping rate and transmission coefficient
derived by using of the reaction flux method. Sec. IV serves as a
summary of our conclusion where some implicate applications of this
study are also discussed.

\section{modulated system-reservoir environment}

The physical scenario depicting the modulation of the bath by an
external noise lives in many different kinds of situations in
forming a non-equilibrium (or non-thermal) system-reservoir coupling
environment \cite{phys1,phys2}. For example, the simple unimolecular
conversion process from $A$ to $B$ in an isomerization reaction is
generally considered to be carried out in a photochemically active
solvent under the influence of external fluctuating light intensity.
In the reaction, fluctuations in the light intensity result in
fluctuations in the polarization of the solvent molecules. Thus the
effective reaction field around the reactant system gets modified.
Given the required stationarity of this non-equilibrium open system
is maintained, the dynamics of barrier crossing evolves amenable to
the present theoretical analysis that follows.

Let us begin our study from the Hamiltonian describing a system of
particles with unit mass bilinearly coupled to a harmonic reservoir
that is modulated by an external noise. Mathematically it reads
\cite{hami}
\begin{eqnarray}
H=\frac{1}{2}p^{2}+U(x)+\sum^{N}_{j=1}\left[\frac{1}{2}(p^{2}_{j}
+\frac{1}{2}\omega_{j}^{2}(q_{j}-c_{j}x)^{2}\right]+H_{\textrm{int}},
\label{eq:hamiltonian}
\end{eqnarray}
where $\{x,p\}$ and $\{q_{j}, p_{j}\}$ are the sets of coordinate
and momentum variables of system and reservoir oscillators,
respectively. $c_{j}x$ measures the interaction between the particle
and reservoir, $U(x)$ is the potential.
$H_{\textrm{int}}=\sum_{j=1}\lambda_{j}q_{j}\epsilon(t)$ represents
the modulating interaction on the reservoir results from the
external noise $\epsilon(t)$ which is assumed to be stationary and
Gaussian with zero mean and decaying second order correlation
function $\langle\epsilon(t)\epsilon(t')\rangle_{e}=2D\psi(t-t')$.
Here $\langle\cdots\rangle_{e}$ implies the averaging over all the
realizations of $\epsilon(t)$ with strength $D$ and $\psi(t)$ is a
relevant memory kernel.

Eliminating the bath variables in the usual way
\cite{met1,met2,met3}, a generalized Langevin equation (GLE) can be
obtained as
\begin{eqnarray}
\ddot{x}+\int^{t}_{0}dt'\gamma(t-t')\dot{x}(t')+\partial_{x}U(x)
=\zeta(t)+\varepsilon(t),\label{eq.GLE}
\end{eqnarray}
where $\zeta(t)=\sum_{j}c_{j}\{[q_{j}(0)-c_{j}x_{0}]
\omega_{j}^{2}\textrm{cos}\omega_{j}t+
p_{j}(0)\omega_{j}\textrm{sin}\omega_{j}t\}$ is the internal
fluctuation force generated from the system-reservoir coupling.
While $\varepsilon(t)=-\int_{0}^{t}\varphi(t-t')\epsilon(t')dt'$
with
$\varphi(t)=\sum_{j}c_{j}\lambda_{j}\omega_{j}\textrm{sin}\omega_{j}t$
is an additional fluctuation force resulted from the external noise
$\epsilon(t)$. The anomalous form of Eq.(\ref{eq.GLE}) suggests that
the system is under the combining government of two forcing
functions. This will in no doubt lead to some novelty results.

Before reaching the central point of this article, let us firstly
digress a little bit about $\zeta(t)$ and $\varepsilon(t)$. Due to
its particular origin, the statistical properties of $\zeta(t)$ are
determined by the initial conditions of the system-reservoir
coupling environment which is assumed to be equilibrium at $t=0$
when the external noise agency has not been switched on. That is
$\langle\zeta(t)\rangle=0$ and
$\langle\zeta(t)\zeta(t')\rangle=k_{B}T\gamma(t-t')$ satisfying the
fluctuation and dissipation theorem \cite{fdt1,fdt2}. However the
statistical properties of $\varepsilon(t)$ may be enslaved to
several aspects of factors such as the normal-mode density of the
bath frequencies, the coupling of the system with the bath, the
coupling of the bath with the external noise and the external noise
itself. The very structure of $\varepsilon(t)$ suggests that this
forcing function is different from a direct driving force acting on
the system. Therefore after the external noise agency is switched
on, the system can be regarded as being driven by an effective noise
$\xi(t)$ $(=\zeta(t)+\varepsilon(t))$ whose correlation is given by
\begin{eqnarray}
\langle\langle\xi(t)\xi(t')\rangle\rangle=k_{B}T\gamma(t-t')+2D\int^{t}_{0}dt_{1}
\varphi(t-t_{1})\int^{t_{1}}_{0}dt_{2}\psi(t_{1}-t_{2})\varphi(t-t_{2}),\label{eq.cor}
\end{eqnarray}
along with $\langle\langle\xi(t)\rangle\rangle=0$, where
$\langle\langle\cdots\rangle\rangle$ means taking two averages
independently. This relation is reminiscent of the familiar
fluctuation-dissipation theorem. However, due to the appearance of
the external noise intensity, it rather serves as a thermodynamic
consistency condition instead.

Due to the Gaussian property of the noises $\zeta(t)$ and
$\epsilon(t)$ and the linearity of the GLE, the joint probability
density function $W(x,v,t; x_{0},v_{0})$ of the system oscillator
must still be written in a Gaussian form \cite{adelm}
\begin{eqnarray}
W(x,v,t;x_{0},v_{0})
=\frac{1}{2\pi|\textbf{A}(t)|^{1/2}}e^{-\frac{1}{2}\left[y^{\dag}(t)\textbf{A}^{-1}(t)y(t)\right]}.\label{eq,pdf}
\end{eqnarray}
where $y(t)$ is the vector $[x-\langle x(t)\rangle,v-\langle
v(t)\rangle]$ and $\textbf{A}(t)$ is the matrix of second moments
with each component as
\begin{subequations}\begin{eqnarray}
A_{11}(t)&=&\langle[x-\langle x(t)\rangle]^{2}\rangle,\\
A_{12}(t)&=&A_{21}(t)=\langle[x-\langle x(t)\rangle][v-\langle
v(t)\rangle]\rangle,\\
A_{22}(t)&=&\langle[v-\langle v(t)\rangle]^{2}\rangle.
\end{eqnarray}\label{A(t)}\end{subequations}
The reduced distribution function can then be yielded by integrating
Eq. (\ref{eq,pdf}) over $v$ as
\begin{eqnarray}
W(x,t; x_{0},v_{0}) =\frac{1}{\sqrt{2\pi
A_{11}(t)}}\textrm{exp}\left[{-\frac{(x-\langle
x(t)\rangle)^{2}}{2A_{11}(t)}}\right].\label{eq,pd}
\end{eqnarray}
in which the average position $\langle x(t)\rangle$ can be obtained
by Laplace solving the GLE. In the case of an inverse harmonic
potential $U(x)=-\frac{1}{2}m\omega^{2}_{b}x^{2}$, it reads
\begin{eqnarray}
\langle
x(t)\rangle&=&\left[1+\omega^{2}_{b}\int^{t}_{0}H(t')dt'\right]x_{0}+H(t)v_{0}
\label{eq,aver}
\end{eqnarray}
where $H(t)$ namely the response function can be yielded from
inverse Laplace transforming
$\hat{H}(s)=[s^{2}+s\hat{\gamma}(s)-\omega^{2}_{b}]^{-1}$ with
residue theorem \cite{ret1,ret2}.

\section{barrier escaping process}

For the activated barrier crossing process, one of the most crucial
factors that we concern is the probability of passing over the
saddle point (namely also the characteristic function) which can
then be determined mathematically by integrating Eq. (\ref{eq,pd})
over $x$ from zero to infinity
\begin{eqnarray}
\chi( x_{0},v_{0};t)&=&\int^{\infty}_{0}W(x,t;
x_{0},v_{0})dx,\nonumber\\
&=&\frac{1}{2}\textrm{erfc}\left(-\frac{\langle
x(t)\rangle}{\sqrt{2A_{11}(t)}}\right),\label{eq,chsi}
\end{eqnarray}
which will lead to a finite real number range from 0 to 1 with 1 for
reactive trajectories and 0 for nonreactive ones. The escape rate of
a particle, defined in the spirit of reactive flux method by
assuming the initial conditions to be at the top of the barrier, can
then be yielded from
\begin{eqnarray}
k(t)&=&\frac{1}{h}\int^{\infty}_{-\infty}dx_{0}
\int^{\infty}_{-\infty}v_{0}
W_{\textrm{st}}(x_{0},v_{0})\delta(x_{0}-x_{b})\chi(
x_{0},v_{0};t)dv_{0}\label{eq,rate}
\end{eqnarray}
in the phase space, where $W_{\textrm{st}}(x_{0},v_{_{0}})$
generally should be an equilibrium Boltzmann distribution that
depends on the initial position and velocity of the particle.
However, in the external noise modulated system-reservoir coupling
environment it should be replaced by a Boltzmann form stationary
probability distribution
$W_{\textrm{st}}(x_{0},v_{0})=\frac{1}{Q}\textrm{exp}
[-\{{\frac{v_{0}^{2}}{2D_{b}}+\frac{\tilde{U}(x_{0})}{D_{b}+\Psi(\infty)}}\}]$
which has been proved in Ref.\cite{dsray3} as a steady-state
solution of the relevant Fokker-Planck equation. In the formula
aforesaid, $Q$ is the partition function,
$\tilde{U}(x)=U_{b}-\frac{1}{2}\Omega_{b}^{2}(x-x_{b})^{2}$ is the
renormalized linear potential near the barrier top with $\Omega_{b}$
an effective frequency and $U_{b}$ the barrier height.
$\Psi(\infty)$ and $D_{b}$ are two asymptotic constants to be
calculated in the long time steady state (refer to Ref.\cite{dsray3}
for detailed information). This stationary distribution for the
non-equilibrium open system is not an equilibrium distribution but
it plays the role of an equilibrium distribution of the closed
system, which may, however, be recovered in the absence of the
external noise.

In general, the above total rate $k(t)$ in Eq.(\ref{eq,rate}) can be
viewed as a generalized TST rate
$k^{\textrm{TST}}=\frac{1}{Qh}e^{-U_{b}/(D_{b}+\Psi(\infty))}$
(where $[D_{b}+\Psi(\infty)]/k_{B}$ in the exponential factor
defines a new effective temperature characteristic of the steady
state of the non-equilibrium open system) multiplied by a factor
between 0 and 1 which describes the possibility of a particle
already escaped from the metastable well to recross the barrier. By
substituting Eqs.(\ref{A(t)}), (\ref{eq,aver}) and (\ref{eq,chsi})
into Eq.(\ref{eq,rate}) we obtain $k(t)=\kappa(t)k^{\textrm{TST}}$
with
\begin{eqnarray}
\kappa(t)&=&\left(1+\frac{A_{11}(t)}{D_{b}H^{2}(t)}\right)^{-1/2},\label{eq,kappa}
\end{eqnarray}
acting as an effective transmission factor which leads immediately
to the Kramers¡¯ formula for the rate constant \cite{kramers,fopm}
in the absence of the external noise.
As expected, both $\kappa(t)$ and $k(t)$ are the functions of the
external noise strength $D$ and the coupling of noise to the bath
modes. The varying of $\kappa(t)$ provides an isolated inspection on
the dynamical corrections of $k(t)$ to the TST rate.

In the following calculations we rescale all the parameters so that
dimensionless quantities such as $k_{B}T=1.0$ are used. We consider
a particular case as an example where the external noise is $\delta$
correlated and the internal is an Ornstein-Uhlenbeck (OU) process,
i.e.,
\begin{subequations}\begin{eqnarray}
&&\langle\epsilon(t)\epsilon(t')\rangle_{e}=2D\delta(t-t'),\\
&&\langle\zeta(t)\zeta(t')\rangle=k_{B}T\frac{g^{2}}{\tau}e^{-|t-t'|/\tau},
\end{eqnarray}\label{pp(t)}\end{subequations}
both symmetric with respect to the time argument and assumed to be
uncorrelated with each other. Where $D$ and $g$ are effective
friction constants and $\tau^{-1}$ is a cutoff frequency of the
system oscillator. It should be noted that for $\tau\rightarrow0$,
the internal noise shown above becomes also $\delta$-correlated.
Asymptotic parameters contained in the stationary probability
distribution $W_{\textrm{st}}(x_{0},v_{_{0}})$ are determined to be
$\Omega_{b}=\omega_{b}$, $D_{b}=k_{B}T$ and $\Psi(\infty)=0$ in the
particular case that we considered here.

\begin{figure}[h]
\centering
\includegraphics[scale=0.7]{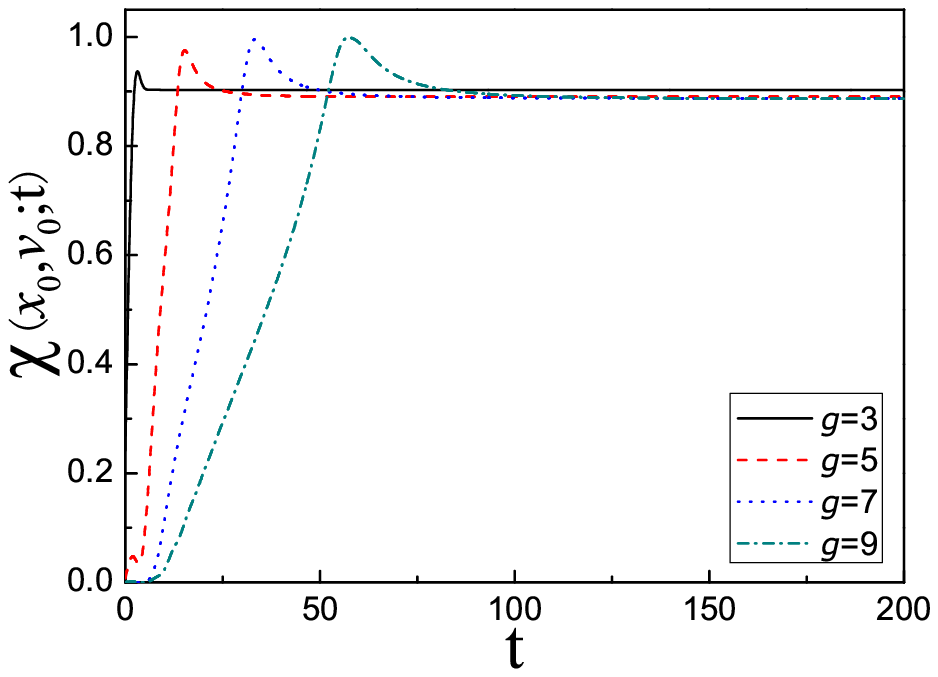}
\includegraphics[scale=0.7]{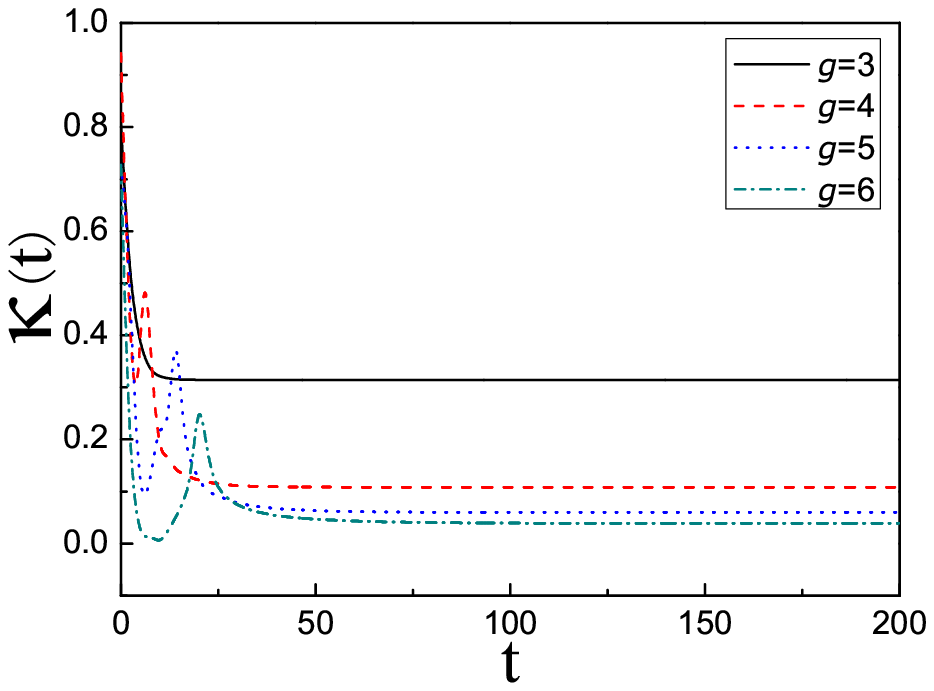}
\caption{(Color online) Instantaneous values of
$\chi(x_{0},v_{0};t)$ and $\kappa(t)$ for various strengths of
purely internal noise ($D=0$). Parameters used are $\tau=3.5$,
$\omega_{b}=1.2$ and $x_{0}=-1.0$, $v_{0}=0.5$
respectively.}\label{Fig1}
\end{figure}
\begin{figure}[h]
\centering
\includegraphics[scale=0.7]{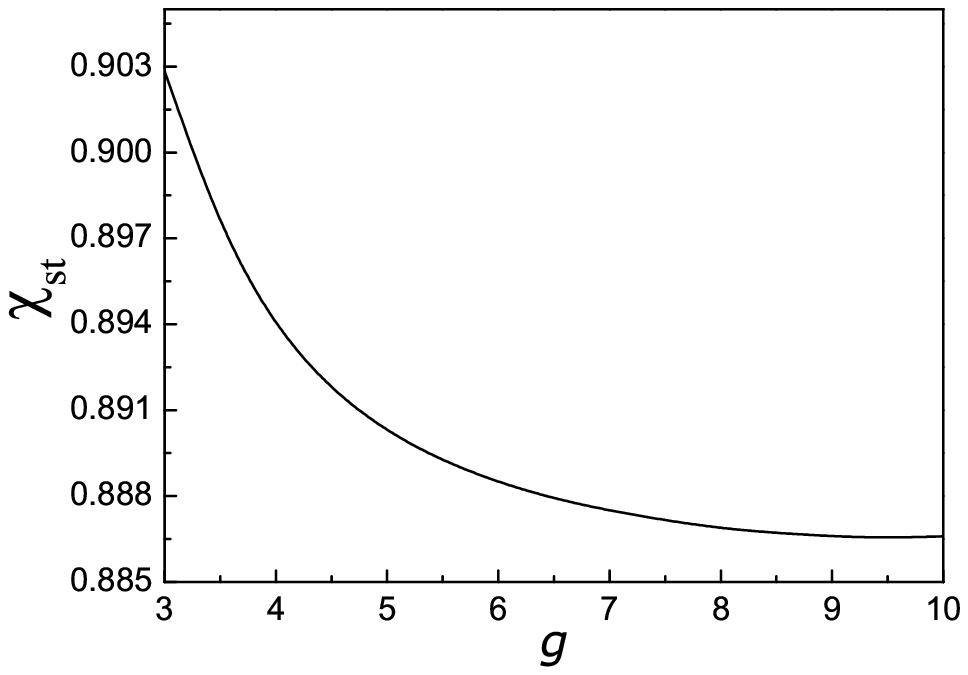}
\includegraphics[scale=0.7]{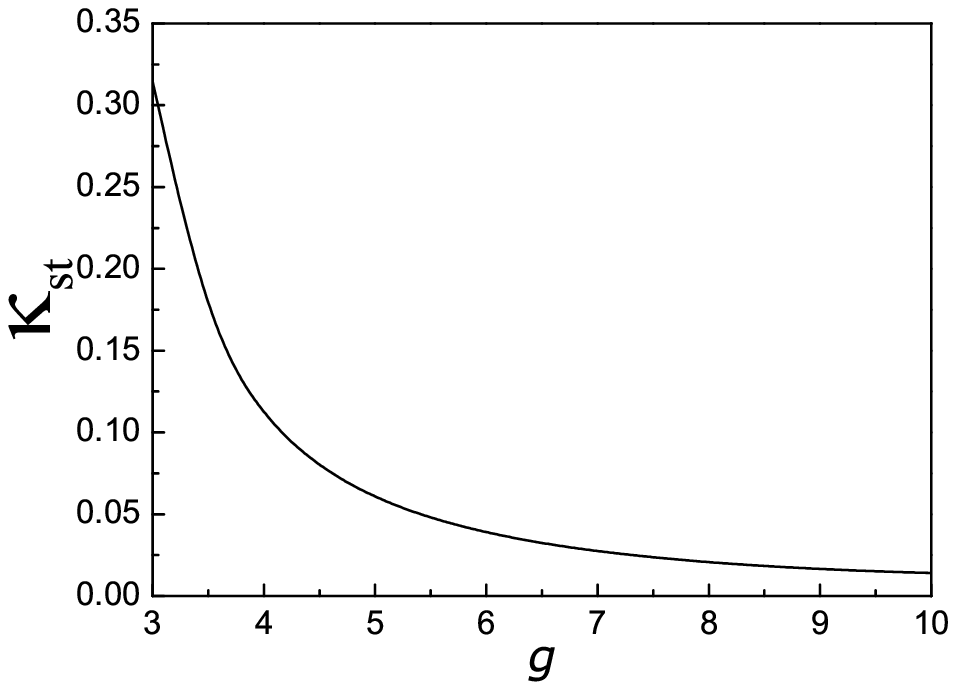}
\caption{Stationary values of $\chi_{\textrm{st}}$ and
$\kappa_{\textrm{st}}$ as a function of the strength of internal
noise $g$ where identical parameters are used as those in
Fig.\ref{Fig1}.}\label{Fig2}
\end{figure}
Firstly, we present a detailed investigation on the rate process
which is purely determined by the internal noise. Fig.\ref{Fig1}
gives the instantaneous values of the barrier passing probability
$\chi(x_{0},v_{0};t)$ and transmission coefficient $\kappa(t)$ for
various strengths of internal noise $g$ where all the system
parameters are set dimensionless and particles are assumed to start
from $x_{0}=-1.0$ position with initial velocity $v_{0}=0.5$. From
which we can see that both $\chi(x_{0},v_{0};t)$ and $\kappa(t)$
evolve asymptotically to a stationary value in the long time limit.
Meanwhile the stationary values of them decrease monotonously as the
increase of the strength of internal noise. This can also be
witnessed in Fig.\ref{Fig2} where the stationary values
$\chi_{\textrm{st}}$ and $\kappa_{\textrm{st}}$ are plotted as a
function of the strength of internal noise $g$. This is a trivial
phenomenon which can be understood in ease for a rate process.
Because a hard dissipative environment is in no doubt harmful to the
diffusing process. Not only will it present a frictional resistance
force but also can it increase the amplitude of fluctuation.
Therefore a smaller and smaller net flux is expected as the
increasing of the strength of the internal noise. Particles which
have already escaped from the potential well will have a big
probability to recross the barrier. Correspondingly related is a
small $\kappa_{\textrm{st}}$ as is shown in Fig.\ref{Fig1}.

\begin{figure}[h]
\centering
\includegraphics[scale=0.7]{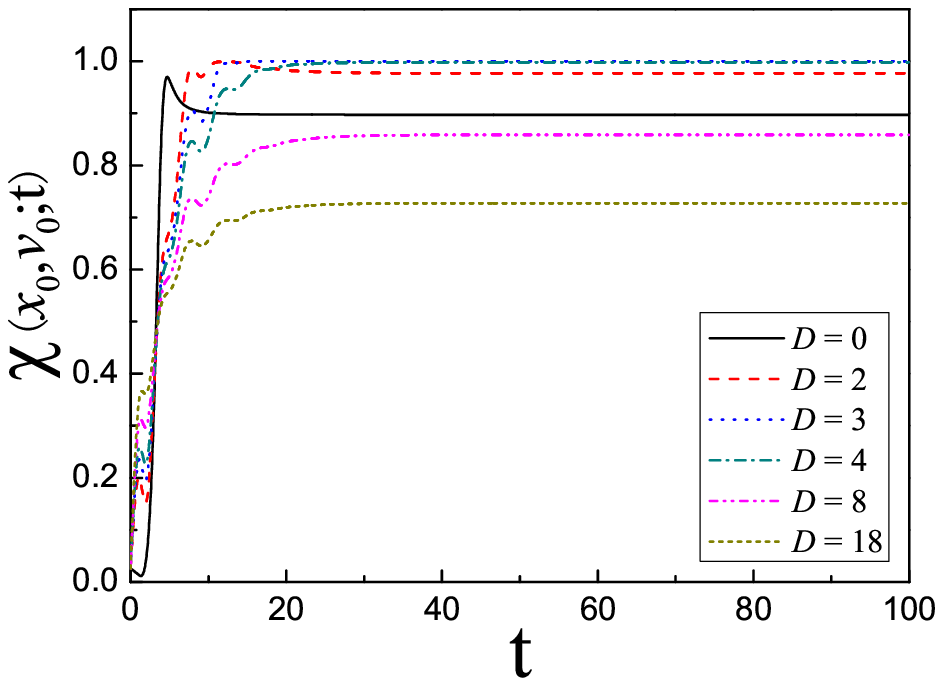}
\includegraphics[scale=0.7]{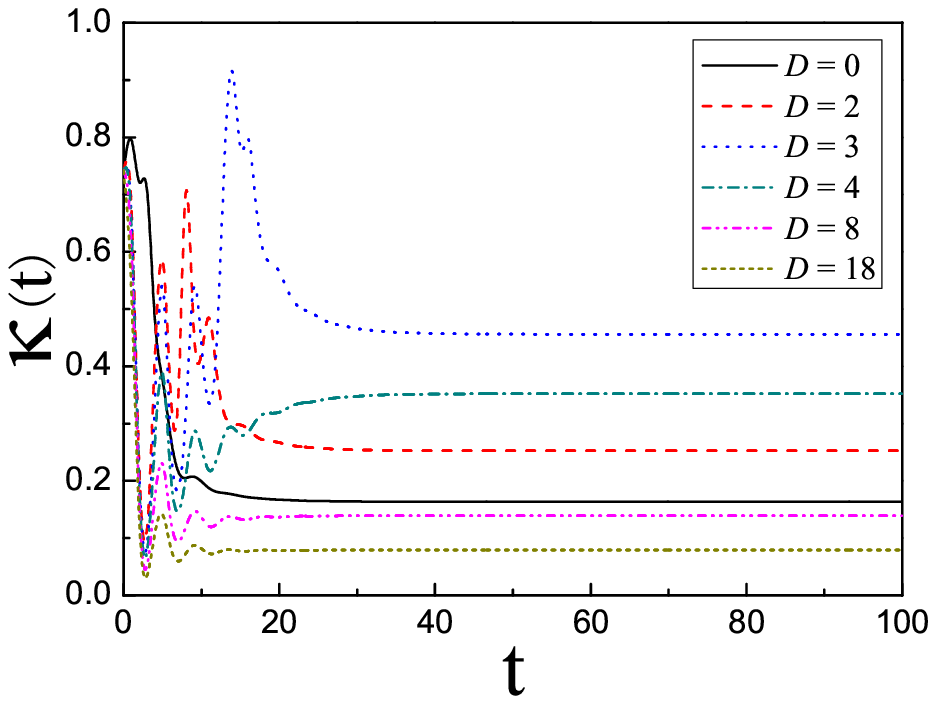}
\caption{(Color online) Instantaneous values of
$\chi(x_{0},v_{0};t)$ and $\kappa(t)$ for various strengths of
external noise. Parameters used are $\tau=3.5$, $\omega_{b}=1.2$,
$g=3.5$ and $x_{0}=-1.0$, $v_{0}=0.5$ for each case.}\label{Fig3}
\end{figure}
\begin{figure}
\centering
\includegraphics[scale=0.7]{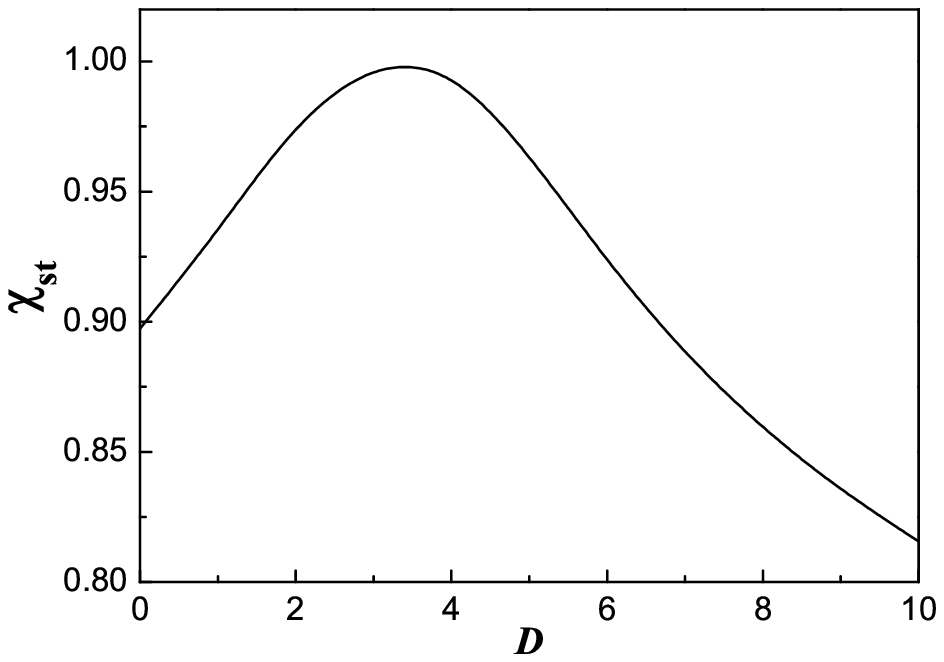}
\includegraphics[scale=0.7]{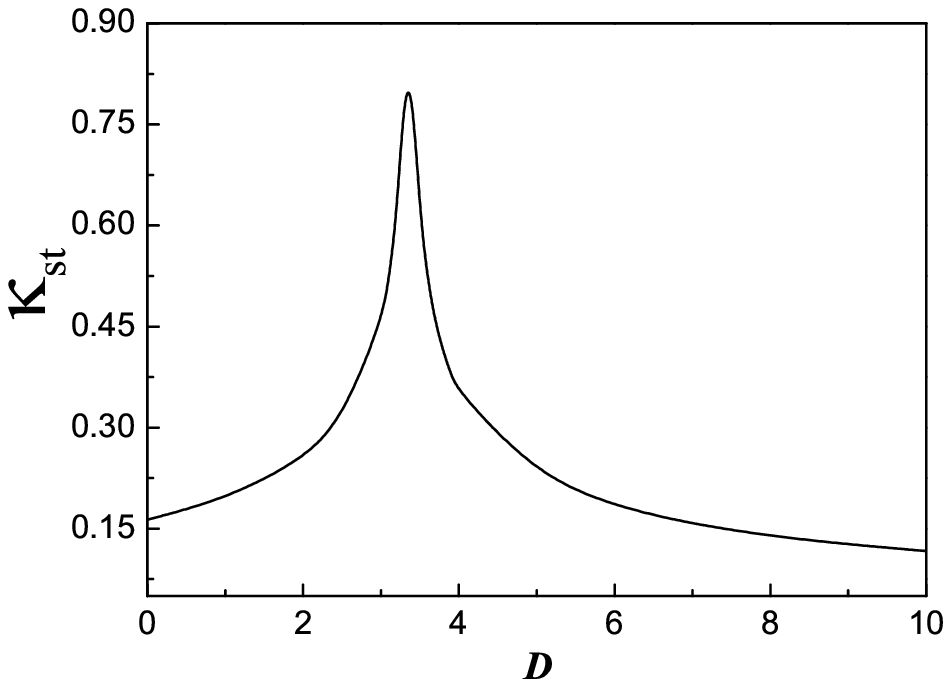}
\caption{Stationary values of $\chi_{\textrm{st}}$ and
$\kappa_{\textrm{st}}$ as a function of the strength of external
noise where identical parameters are used as those in
Fig.\ref{Fig3}.}\label{Fig4}
\end{figure}

In further, let us turn to the most important point of our study,
i.e., what will happen if the rate process is additionally modulated
by an external noise? In order to obtain a explicit elucidation of
this problem we present in Fig.\ref{Fig3} again the instantaneous
values of $\chi(x_{0},v_{0};t)$ and $\kappa(t)$ at various strengths
of external noise where identical system parameters are used as
those in Fig.\ref{Fig1} except for $D=0$, $2$, $3$, $4$, $8\textrm{
and }18$ respectively. The stationary values of them
($\chi_{\textrm{st}}$ and $\kappa_{\textrm{st}}$) are also exhibited
as a function of the external noise strength $D$ in Fig.\ref{Fig4}
in the meantime. From which we can see a non-trivial phenomenon that
both $\chi_{\textrm{st}}$ and $\kappa_{\textrm{st}}$ varies
non-monotonously as the increasing of the strength of external
noise. That is to say not all the cases of external noise modulating
is harmful to the rate process. Sometimes it is even beneficial to
the diffusing of the particle. For it can result in a bigger
$\chi_{\textrm{st}}$ and $\kappa_{\textrm{st}}$ comparing to the
pure internal case. This reveals, in the combining effect of
internal and external noises, the particle will be able to get a
biggest probability to escape from the potential well. Therefore a
notable rate of net flux is expected. In further, we can infer from
it that there lives an optimal strength of external noise modulation
for the particle to obtain a biggest barrier escaping probability.
In the particular case that is considered here this optimal value is
about $D=3.4$ (dimensionless) as can be calculated following the
procedures in the context foreknown.

\section{SUMMARY and discussion}

In summary, we have studied in this paper the time-dependent barrier
passage of an external noise modulated system-reservoir coupling
environment. We proposed analytically the barrier passing
probability and the rate of barrier escaping. The influence of the
external noise on the dynamical barrier escaping process is
investigated carefully by calculating the effective transmission
coefficient of an external $\delta$-correlated noise modulated
internal Ornstein-Uhlenbeck process for an example. The main
conclusions of our study is that not all the cases of external noise
modulating is harmful to the rate process. Sometimes it is even
beneficial to the diffusion of the particle. There exists an optimal
strength of external noise modulation for the particle to obtain a
biggest probability to escape from the potential well to form a
notable rate of effective net flux.

In applications and industrial processing, the creation of a typical
non-equilibrium open situation by modulating a bath with the help of
an external noise is not an uncommon phenomenon. The external agency
generating noise does work on the bath by stirring, pumping,
agitating, etc., to which the system dissipates internally. However,
provided the long-time limit of the moments for the stochastic
processes pertaining to the external and internal noises
characterized by arbitrary decaying correlation functions exist, the
expression for the effective transmission coefficient of barrier
crossing rate for the open system we derive here is fairly general.
Therefore, we believe that these considerations are likely to be
important in other related issues in non-equilibrium open systems
and may serve as a basis for studying processes occurring within
irreversibly driven environments \cite{jray,rher} and for thermal
ratchet problems \cite{rdas}. The externally generated
non-equilibrium fluctuations can bias the Brownian motion of a
particle in an anisotropic medium and may also be used for designing
molecular motors and pumps.

\section * {ACKNOWLEDGEMENTS}

This work was supported by the Shandong Province Science Foundation
for Youths (Grant No.ZR2011AQ016) and the  Shandong Province
Postdoctoral Innovation Program Foundation (Grant No.201002015).

\end{document}